\documentstyle[12pt,aps,epsf,floats]{revtex}
\draft

\begin{document}

\title{Evolution of the pairing pseudogap in the spectral function with
interplane anisotropy}
\author{G. Preosti$^{1}$, Y. M. Vilk$^{2}$, and M. R. Norman$^{1}$}
\address{$(1)$
Materials Sciences Division, Argonne National Laboratory, Argonne, IL 60439}
\address{$(2)$
2100 Valencia Dr. Apt. 406, Northbrook, IL 60062}
\date{\today}
\maketitle

\begin{abstract}
We study the pairing pseudogap in the spectral function as a function of
interplane coupling. The analytical expressions for the self-energy in the
critical regime are obtained for any degree of anisotropy. The frequency
dependence of $\Sigma \left( \omega \right) $ is found to be qualitatively
different in two and three dimensions, and the crossover from two to three
dimensional behavior is discussed. In particular, by considering the
anisotropy of the Fermi velocity and gap along the Fermi surface, we can
qualitatively explain recent photoemission experiments on high temperature
superconductors concerning the temperature dependent Fermi arcs seen in the
pseudogap phase.
\end{abstract}

\section{Introduction}

\smallskip

It is clearly shown by experimental data that underdoped cuprates exhibit a
large suppression of low energy spectral weight at temperatures higher than
the superconducting (SC) critical temperature, which constitutes the so
called pseudogap phenomenon. Early evidence came from a number of probes,
such as NMR, infrared reflectivity, and specific heat\cite{Randeira}, while
recent angle-resolved photoemission (ARPES) experiments have added
qualitatively new information, and attracted considerable attention to this
phenomenon\cite{ARPES,Mike1}. The picture that emerges from ARPES data
is that of a highly anisotropic pseudogap, for $T_{c}<T<T^{*}$, which in
magnitude and angular dependence resembles the superconducting gap below $%
T_{c}$ and smoothly evolves into it as the temperature is lowered. With
increasing underdoping, the interplane anisotropy of these materials
increases\cite{PG}, $T_{c}$ goes down, and the amplitude of the gap and the
temperature $T^{*}$, at which the pseudogap opens, increases, thus widening
the extent of the pseudogap phase. Moreover the temperature $T^{*}$, and the
strength of the pseudogap feature, are strongly dependent on the region of
the Fermi surface that is probed, being quite enhanced close to the $\left(
\pi ,0\right) $ point and vanishing along the diagonal direction
in the Brillouin zone.

These measurements have spawned a number of theoretical interpretations, and
no consensus has yet being reached. Theoretical models span a wide range of
possible mechanisms, including RVB-like pairing of chargeless spinons\cite
{RVB}, antiferromagnetic correlations\cite{magnetic}, and preformed pairs%
\cite{preformed}. Here we will concentrate on the effects of dimensionality%
\cite{Qijin} and intra-plane anisotropy on the properties of the pseudogap.

It is the purpose of this work to show under what general conditions an
observable pseudogap can be expected as a precursor of superconductivity in
the weak to intermediate coupling regime, and to demonstrate the crucial
role of both in plane and interplane anisotropy in determining the
properties of the pseudogap. We adopt a procedure for obtaining an
analytical estimate of the electron self energy in the renormalized
classical regime of the pairing fluctuations, which is similar to the one
adopted by Vilk et. al.\cite{Vilk,VilkSolo}. This procedure is
justifiable up to intermediate coupling.

First we find that a very weak pseudogap exists in the spectral function $%
A\left( \omega ,\mathbf{k}\right) $ even in a perfectly isotropic three
dimensional (3D) superconductor but it is of a very different nature from
the one found in exactly two spatial dimensions, and even very close to the
transition temperature it would be practically unobservable at present day
resolution. By contrast, for highly anisotropic materials, the size of the
pseudogap well above the SC critical temperature becomes almost the same as 
in the ordered state.
Therefore, unless strong coupling is invoked, a high
interplane anisotropy is required to make the pseudogap observable. The
condition for having a strong quasi-2D pseudogap effect in $A\left( \omega ,%
\mathbf{k}\right) $ can be expressed as: $\Delta _{\mathbf{k}}\left(
0\right) \gtrsim $ $\omega _{X}$, 
where $\Delta _{\mathbf{k}}\left( 0\right) 
$ is the k-dependent value of the zero temperature gap amplitude, and $%
\omega _{X}=$ $\left( \xi _{z}/\xi _{xy}\right) 
v_{F}\left( \mathbf{k}\right) q_{BZ}^{\left( z\right) } $ is a 
dimensional crossover frequency which
involves both the interplane anisotropy through the pairing correlation
length anisotropy ratio $\xi _{z}/\xi _{xy}$, and the in plane anisotropy of
the Fermi velocity $v_{F}\left( \mathbf{k}\right) $ (here $q_{BZ}^{\left(
z\right) }$ is the Brillouin zone boundary in the direction of the
anisotropy axis). Second, we find that all the properties of the pseudogap,
and in particular its temperature evolution, are strongly affected by the
ratio $\Delta _{\mathbf{k}}\left( 0\right) /v_{F}\left( \mathbf{k}\right) $,
which depends on the particular point probed on the Fermi surface. We find
that this dependence on the in plane anisotropy naturally accounts for the
ARPES observation of Fermi arcs in the pseudogap regime, and also the
apparently different line shape and temperature dependence of
the pseudogap at different points on the Fermi surface. In particular, we
explain the experimental observation that the pseudogap seems to fill up
close to the $\left( \pi ,0\right) $\ point, and to close in for points
close to the zone diagonal.

\section{Formalism}

We evaluate the effects of pairing fluctuations on the electronic
self-energy of anisotropic superconductors in the one loop approximation.
The expression for the self energy has the form:

\begin{equation}
\Sigma \left( \omega ,\mathbf{k}\right) =g_{\mathbf{k}}^{2}\int \frac{%
d^{3}q\,d\omega ^{\prime }}{2\pi \Omega }\frac{\chi ^{\prime \prime }\left(
\omega ^{\prime },\mathbf{q}\right) \left\{ \coth \left( \frac{\omega
^{\prime }}{2T}\right) -\tanh \left( \frac{\widetilde{\varepsilon }\left( -%
\mathbf{k+q}\right) }{2T}\right) \right\} }{\omega -\omega ^{\prime }+%
\widetilde{\varepsilon }\left( -\mathbf{k+q}\right) +i0^{+}}
\label{GeneralSelfEnergy}
\end{equation}

\noindent where $g_{\mathbf{k}}^{2}$ is the effective coupling 
constant\cite{gtilda}
between electrons and pairing fluctuations, $\Omega $ the volume
of the reciprocal lattice primitive cell, 
$\widetilde{\varepsilon }\left( \mathbf{k}\right) $ the
energy dispersion relative to the chemical potential, and $\chi \left(
\omega ,\mathbf{q}\right) $ the pairing susceptibility. In this
formula and what follows we have set the in plane lattice constant $a$ as
the unit of length, the interplane lattice constant as $a_{z}=\pi
/q_{BZ}^{\left( z\right) }$, and we also consider natural units: $\hbar
=k_{B}=1$.

\strut In the critical regime close to the SC instability, the pairing
susceptibility $\chi $ is strongly peaked at $q=0$, $\omega =0$. It
is this peak in the pairing susceptibility that gives rise to the pseudogap
effect. For the purpose of this work, it will be sufficient to assume that
for $T<T_{X}$, where $T_{X}$ is the crossover temperature to the critical
regime, $\chi $ can be approximated in the neighborhood of $\left( \omega ,%
\mathbf{q}\right) =0$ by the following Lorentzian form:

\begin{equation}
\chi \left( \omega ,\mathbf{q}\right) =\frac{\chi _{0}\left(
0,0,T_{X}\right) }{\xi _{0}^{2}}\frac{1}{\xi ^{-2}-\frac{i\omega }{D}+\left(
q_{x}^{2}+q_{y}^{2}\right) +\gamma ^{2}q_{z}^{2}}
\label{GeneralSusceptibility}
\end{equation}

\noindent where $\chi _{0}\left( 0,0,T_{X}\right) $ is the pairing
susceptibility for non-interacting electrons at the crossover temperature, $%
\xi _{0}=\xi _{0xy}$ is the in plane coherence length, and $D$ the
microscopic diffusion constant. Two other fundamental quantities appear in
this expression: the in plane correlation length $\xi =\xi _{xy}$ which
diverges at the transition temperature, and the anisotropy parameter $\gamma
=\xi _{0z}/\xi _{0}$ which expresses the interplane anisotropy. In the
pairing channel, the expansion for $\chi ^{-1}\left( \omega ,\mathbf{q}%
\right) $ can also contain a real linearly dependent frequency term, $\omega
/D_{1}$. However, it can be shown by direct evaluation of the bare
susceptibility $\chi _{0}\left( \omega ,\mathbf{q}\right) =-\left[ \frac{1}{%
\beta }\sum_{ik_{n}}\sum_{\mathbf{k}}G^{0}\left( ik_{n},\mathbf{k}\right)
G^{0}\left( iq_{n}-ik_{n},\mathbf{q-k}\right) \right] _{iq_{n\rightarrow
\omega +i\delta }}$ that at the beginning of the critical regime $D=\left( 
\frac{-1}{\xi _{0} ^{2}\chi _{0}}\mbox{Im}\left[ \frac{\partial \chi _{0}\left(
q,0\right) }{\partial \omega }\right] \right) ^{-1}\sim T$, while $%
D_{1}=\left( \frac{1}{\xi _{0} ^2\chi _{0}}\mbox{Re}\left[ \frac{\partial \chi
_{0}\left( q,0\right) }{\partial \omega }\right] \right) ^{-1}\sim E_{F}$,
and therefore the term $\omega /D_{1}$ should be negligible with respect to
the $i\omega /D$ term in a degenerate electron system. Since ARPES data show
a large Fermi surface\cite{FermiS} we neglect this additional term. The case
in which the $\omega /D_{1}$ term is dominant has being considered in
references\cite{Qijin,Maly}.

\smallskip

In mean field theory, the correlation length $\xi $ diverges as $1/\sqrt{%
T-T_{c}}$. However, mean field theory violates the Mermin-Wagner theorem%
\cite{MW}, and therefore cannot give a viable description of the effects of
pairing fluctuations in low dimensional systems. In order to go beyond mean
field theory we will utilize the fluctuation-dissipation theorem:

\begin{equation}
\int \frac{d^{3}q\,}{\Omega }\int \frac{d\omega }{2\pi }\chi ^{\prime \prime
}\left( \omega ,\mathbf{q}\right) \coth \frac{\omega }{2T}=\frac{1}{2}\left(
\left\langle \widetilde{\Delta }\widetilde{\Delta }^{+}\right\rangle
+\left\langle \widetilde{\Delta }^{+}\widetilde{\Delta }\right\rangle
\right)   \label{FluctuationDissipationTheorem}
\end{equation}
where $\widetilde{\Delta }=\sum_{j}\eta _{ij}c_{i\uparrow }c_{j\downarrow }$
is the local pairing\cite{symmetry} operator and $\left\langle
...\right\rangle $ indicates thermodynamic averaging. In order to satisfy
the Mermin-Wagner theorem, the right hand side of eq. (\ref
{FluctuationDissipationTheorem}) must remain finite (the mean field
assumption leads to a divergence in 2D\cite{Vilk}). Moreover, the phase
fluctuation scenario consists in having a state with a non-zero local
superconducting pairing amplitude, but no off diagonal long range order due
to thermal fluctuations in the phase of the order parameter. It is important
to realize that the locally ordered part of the pairing amplitude cannot be
simply identified with the full correlator in equation (\ref
{FluctuationDissipationTheorem}) because the latter quantity is large and
nonzero even in the non-interacting system. In order to make the connection
between the local pairing amplitude in the pseudogap regime and the integral
of $\chi ^{\prime \prime }\left( \omega ,\mathbf{q}\right) $, we first note
that Monte Carlo simulations\cite{Singer} show that the temperature
dependence of the quantity $\frac{1}{2}\left( \left\langle \widetilde{%
\Delta }\widetilde{\Delta }^{+}\right\rangle +\left\langle \widetilde{\Delta 
}^{+}\widetilde{\Delta }\right\rangle \right) $ goes through a minimum at
some crossover temperature $T_{X}$, and below $T_{X}$ it increases as if it
were in the ordered state. Clearly this growth of the $\frac{1}{2}\left(
\left\langle \widetilde{\Delta }\widetilde{\Delta }^{+}\right\rangle
+\left\langle \widetilde{\Delta }^{+}\widetilde{\Delta }\right\rangle
\right) $ correlator with decreasing temperature is due to the contribution
of thermal critical fluctuations. We thus identify the local
superconducting pairing amplitude with the almost singular contribution of
the classical fluctuations to the correlator $\frac{1}{2}\left( \left\langle 
\widetilde{\Delta }\widetilde{\Delta }^{+}\right\rangle +\left\langle 
\widetilde{\Delta }^{+}\widetilde{\Delta }\right\rangle \right) $: 
\begin{equation}
\int \frac{d^{3}q\,}{\Omega }\int \frac{d\omega }{2\pi }\chi ^{\prime \prime
}\left( \omega ,\mathbf{q}\right) \left[ \coth \frac{\omega }{2T}-\text{sign}%
\left( \omega \right) \right] =\widetilde{\Delta }_{cl}^{2}\left( T\right)
\qquad T<T_{X}  \label{ClassicalFluctuations}
\end{equation}
where we will assume that $\widetilde{\Delta }_{cl}\left( T\right) $ has the
same temperature dependence as the BCS superconducting gap function and
vanishes at the mean field critical temperature $T_{c}^{MF}$. More
specifically, as it will be clear later from the expression for the
self-energy in the pseudogap regime, $\widetilde{\Delta }_{cl}^{2}\left(
T\right) =\Delta _{\mathbf{k}}^{2}\left( T\right) /g_{\mathbf{k}}^{2}$. We
also note that having the crossover temperature to the pseudogap regime of
the order of $T_{c}^{MF}$ is consistent with \emph{\ }recent data\cite
{Giap} which show $T^{*} \sim T_{c}^{MF}$. Since the largest contribution
in the frequency integral comes from $\omega \ll T$ in the critical regime ($%
T<T_{X}$), eq. (\ref{ClassicalFluctuations}) can be approximated by:

\begin{equation}
T\int_{0}^{\overline{q}_{c}}\frac{d^{3}q\,}{\Omega }\chi \left( 0,\mathbf{q}%
\right) =\widetilde{\Delta }_{cl}^{2}\left( T\right)  \label{DeltaCl}
\end{equation}
Here $\overline{q}_{c}$ is a momentum space cutoff. In the in plane
direction, we set 
$\overline{q}_{c}^{\left( xy\right) }=q_{c}=\pi /\xi _{0}$. In the
out of plane direction, the quantity $\pi /\xi _{0z}=\pi /\gamma \xi _{0}$
can become larger than the Brillouin zone boundary $q_{BZ}^{\left( z\right)
} $ for sufficiently small $\gamma $, and hence we chose 
$\overline{q}_{c}^{\left( z\right) }=q_{c}^{\left( z\right) }$ to be
\begin{equation}
q_{c}^{\left( z\right) }=\min \left[ q_{c}/\gamma ,q_{BZ}^{\left( z\right)
}\right] \text{.}  \label{qcz}
\end{equation}
We checked that the exact value of the cutoff does not significantly affect
the results.

With the help of equations (\ref{GeneralSusceptibility}) and (\ref{DeltaCl}%
), we can obtain the temperature dependence of the correlation length $\xi $%
\cite{CutoffProcedure}:

\begin{equation}
\frac{q_{c}^{\left( z\right) }}{q_{BZ}^{\left( z\right) }}\left\{ \left(
1+\left( q_{c}\xi \right) ^{-2}\right) \left( \frac{\sqrt{1+\left( q_{c}\xi
\right) ^{-2}}}{\sqrt{1-\widetilde{\gamma }^{2}}}\text{atanh}\frac{\sqrt{1-%
\widetilde{\gamma }^{2}}}{\sqrt{1+\left( q_{c}\xi \right) ^{-2}}}-\frac{%
\text{atan}\left( \widetilde{\gamma }q_{c}\xi \right) }{\widetilde{\gamma }%
q_{c}\xi }\right) -\frac{1}{3}\right\} =\frac{T_{0}}{T}\frac{\widetilde{%
\Delta }_{cl}^{2}\left( T\right) }{\widetilde{\Delta }_{cl}^{2}\left(
0\right) }  \label{CorrelationLength}
\end{equation}
where $\widetilde{\gamma }=\gamma q_{c}^{\left( z\right) }/q_{c}$. The
parameter $T_{0}$ is defined by:

\[
T_{0}=2\pi \xi _{0}^{2}\widetilde{\Delta }_{cl}^{2}\left( 0\right) /\chi
_{0} 
\]
The meaning of $T_{0}$ is particularly clear in two dimensions, in which
from eq. (\ref{CorrelationLength}) $\xi \propto e^{\frac{T_{0}}{T}}$. In the
weak coupling limit, $\xi _{0}\sim \frac{v_{F}}{\Delta }$ and $T_{0}$ is
proportional to the Fermi energy and therefore is very large compared to $%
T_{c}^{MF}$. As we shall see below, this implies an extremely narrow
fluctuation regime unless $\gamma $ is practically zero. On the other hand,
in the intermediate coupling regime we can expect $\xi _{0}\sim 1$, which
leads to a sufficiently large fluctuation regime. In the rest of this work
we will choose $\xi _{0}\sim 1$, and $T_{0}\sim \Delta _{\max }\left(
0\right) $ unless otherwise stated (in particular all Figures have being
generated by choosing $\xi _{0}=2.0$, and $T_{0}=\Delta _{\max }\left(
0\right) $).

From equation (\ref{CorrelationLength}) we can find the superconducting
transition temperature, by letting $\xi \rightarrow \infty $:

\begin{equation}
T_{c}\left( \gamma \right) =\frac{q_{BZ}^{\left( z\right) }}{q_{c}^{\left(
z\right) }}\left[ \frac{\text{atanh}\left( \sqrt{1-\widetilde{\gamma }^{2}}%
\right) }{\sqrt{1-\widetilde{\gamma }^{2}}}-\frac{1}{3}\right] ^{-1}\frac{%
\Delta ^{2}\left( T_{c}\right) }{\Delta ^{2}\left( 0\right) }T_{0}
\label{CriticalTemperature}
\end{equation}
This is still an implicit equation for $T_{c}$, but it clearly shows two
important points. First, $T_{c}\left( \gamma \right) $ is a monotonically
increasing function of $\gamma $, and as the interplane anisotropy increases
($\gamma \rightarrow 0$), the critical temperature approaches zero
logarithmically in $\gamma $, therefore satisfying the Mermin-Wagner theorem. 
Clearly
in this limit, a Kosterlitz-Thouless (KT) transition may set in at some
finite temperature, but we will not consider this possibility here and we
will assume that the SC critical temperature due to 3D effects is higher
then $T_{c}^{KT}$. Second, for any finite value of the anisotropy parameter $%
\gamma $, $T_{c}$ is a monotonically increasing function of $T_{0}$,
saturating at $T_{c}^{MF}$ for $T_{0}\gg T_{c}^{MF}$. That is, in the weak
coupling limit, the temperature range of the pseudogap phase is very small 
even for
quasi 2D materials, while for intermediate coupling, the temperature range 
of the
pseudogap phase can be appreciable.

We are now in a position to estimate the imaginary part of the self energy.
Close to the SC transition temperature, classical fluctuations give a
dominant contribution to the self energy. We can therefore approximate eq. (%
\ref{GeneralSelfEnergy}) with its classical part $\Sigma \approx \Sigma
_{cl} $. Using (\ref{GeneralSelfEnergy}), (\ref{GeneralSusceptibility}), and 
\cite{CutoffProcedure} the imaginary part of the self energy $\Sigma
_{cl}^{\prime \prime }$ can be evaluated analytically:

\begin{eqnarray}
\Sigma ^{\prime \prime }\left( \omega ,{\mathbf k},\gamma ,T\right) &=&-\frac{%
\pi }{2}\frac{q_{c}^{\left( z\right) }}{q_{BZ}^{\left( z\right) }}\frac{%
\Delta _{\mathbf{k}}^{2}\left( 0\right) }{v_{\mathbf{k}}q_{c}}\frac{T}{T_{0}}%
\theta \left( v_{\mathbf{k}}q_{c}-\left| \omega \right| \right) \left[ -%
\frac{1}{2}\left( 1-\left( \frac{\omega }{v_{\mathbf{k}}q_{c}}\right)
^{2}\right) +\left( 1+\left( q_{c}\xi \right) ^{-2}\right) \times \right.
\label{GeneralSigIm} \\
&&\left. \frac{1}{\widetilde{\gamma }}\ln \frac{\widetilde{\gamma }\sqrt{%
1+\left( q_{c}\xi \right) ^{-2}}+\sqrt{\widetilde{\gamma }^{2}+\left(
q_{c}\xi \right) ^{-2}+\left( 1-\widetilde{\gamma }^{2}\right) \left( \frac{%
\omega }{v_{\mathbf{k}}q_{c}}\right) ^{2}}}{\left( 1+\widetilde{\gamma }%
\right) \sqrt{\left( q_{c}\xi \right) ^{-2}+\left( \frac{\omega }{v_{\mathbf{%
k}}q_{c}}\right) ^{2}}}\right]  \nonumber
\end{eqnarray}
In this expression we have taken into consideration only the most important
contribution coming from the pole in $\chi $ at $i\omega =D\left( \xi
^{-2}+q_{x}^{2}+q_{y}^{2}+\gamma ^{2}q_{z}^{2}\right) \ll T$. Moreover, we
expanded the single particle energy up to first order in $\mathbf{q}$ ($%
\epsilon \left( \mathbf{q}-\mathbf{k}\right) \simeq \epsilon \left( -\mathbf{%
k}\right) +\mathbf{v}_{-\mathbf{k}}\cdot \mathbf{q}$, where $\mathbf{k}$ is
on the Fermi surface and we use inversion symmetry to set $\mathbf{v}_{-%
\mathbf{k}}=-\mathbf{v}_{\mathbf{k}}$ ), and integrated in momentum space
using the cutoff $\overline{q}_{c}$. Notice that this momentum integration
is essential in order to account for the effect of dimensionality. Indeed,
neglecting the $\mathbf{q}$ dependence in the Green`s function in eq. (\ref
{GeneralSelfEnergy}) would lead to a pseudogap in all dimensions. The factor 
$\theta \left( v_{\mathbf{k}}q_{c}-\left| \omega \right| \right) $ also
comes from this momentum integration and introduces a convenient frequency
cutoff. The real part of the self energy $\Sigma ^{\prime }$ can be obtained
numerically for arbitrary $\gamma $ by using the Kramers-Kronig relations.
Notice that in the strictly two dimensional case ($\gamma =0$), the real and
imaginary part of the self energy can both be expressed analytically\cite
{Vilk}.

\smallskip

\section{T=T$_{c}\left( \gamma \right) $:}

\smallskip

We will start our study of the effect of dimensionality by analyzing the
spectral function when the pseudogap effect is the most noticeable, that is
at $T=T_{c}\left( \gamma \right) $ where the correlation length $\xi =\infty 
$. At exactly the phase transition the self-energy is singular, and this
leads to a pseudogap at $T=T_{c}$. However, as we will see below, the
qualitative shape of the spectral function and the strength of the pseudogap
differ dramatically in 2D and 3D.

\begin{figure}
\centerline{
\epsfxsize=4.0in
\epsfbox{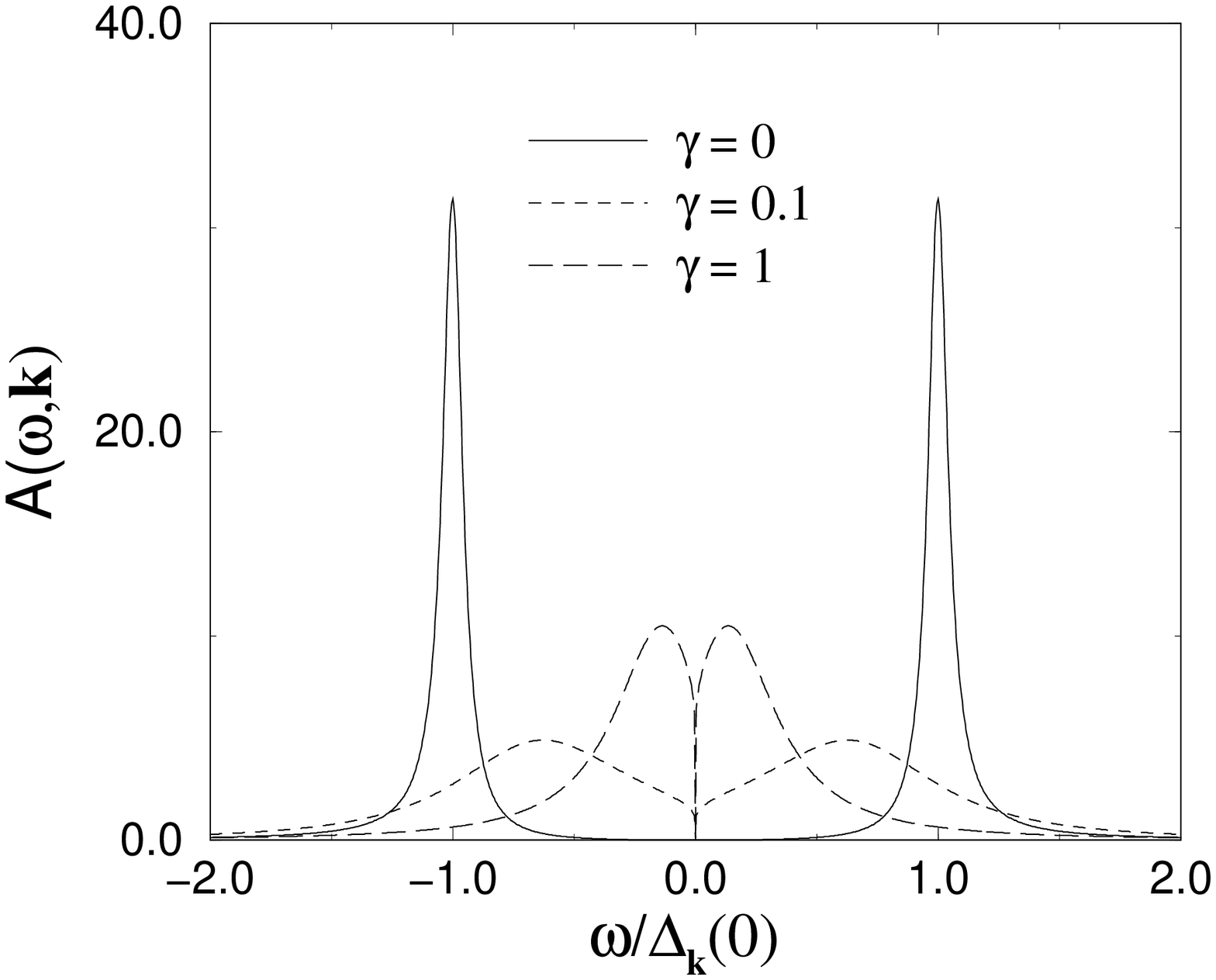}
}
\vspace{0.5cm}
\caption{
Plot of the spectral function
\protect\cite{broadening} $A\left( \omega ,{\mathbf k},\gamma ,T=T_{c}\left(
\gamma \right) \right) $ as a function of $\omega $ for three different
values of the anisotropy parameter $\gamma $, and the same value of the
ratio $\Delta \left( \mathbf{k}\right) /\pi v_{\mathbf{k}}=0.064$. Each spectral
function is plotted at the transition temperature corresponding to $\gamma $.
}
\label{sf_vf=5_g=0.01.1_t0=1}
\end{figure}

In Fig.(\ref{sf_vf=5_g=0.01.1_t0=1}) the spectral function $A\left( \omega ,%
{\mathbf k},\gamma ,T=T_{c}\left( \gamma \right) \right) $ is plotted for
three distinct values of the anisotropy parameter ($\gamma =0,0.1,1$), with
each curve plotted at the respective SC transition temperature $T_{c}\left(
\gamma \right) $, all other parameters being the same. The spectral function
in 2D, $A\left( \omega ,{\mathbf k},0,T=T_{c}\left( 0\right)=0 \right) $, is
simply given by two sharp peaks located at $\omega =\pm \Delta _{\mathbf{k}%
}\left( 0\right) $\cite{broadening}. On the other hand, in the isotropic 3D
case ($\gamma =1$), the spectral function consists of two broad maxima very
close to each other; although $A\left( \omega ,{\mathbf k},\gamma
=1,T=T_{c}\left( 1\right) \right) $ vanishes right at $\omega =0$ because of
the singularity in $\Sigma ^{\prime \prime }$, it is small only in an
extremely narrow range of frequencies. As suggested by the $\gamma =0.1$
curve, a fairly large anisotropy is required to obtain two well separated
peaks.

\begin{figure}
\centerline{
\epsfxsize=4.0in
\epsfbox{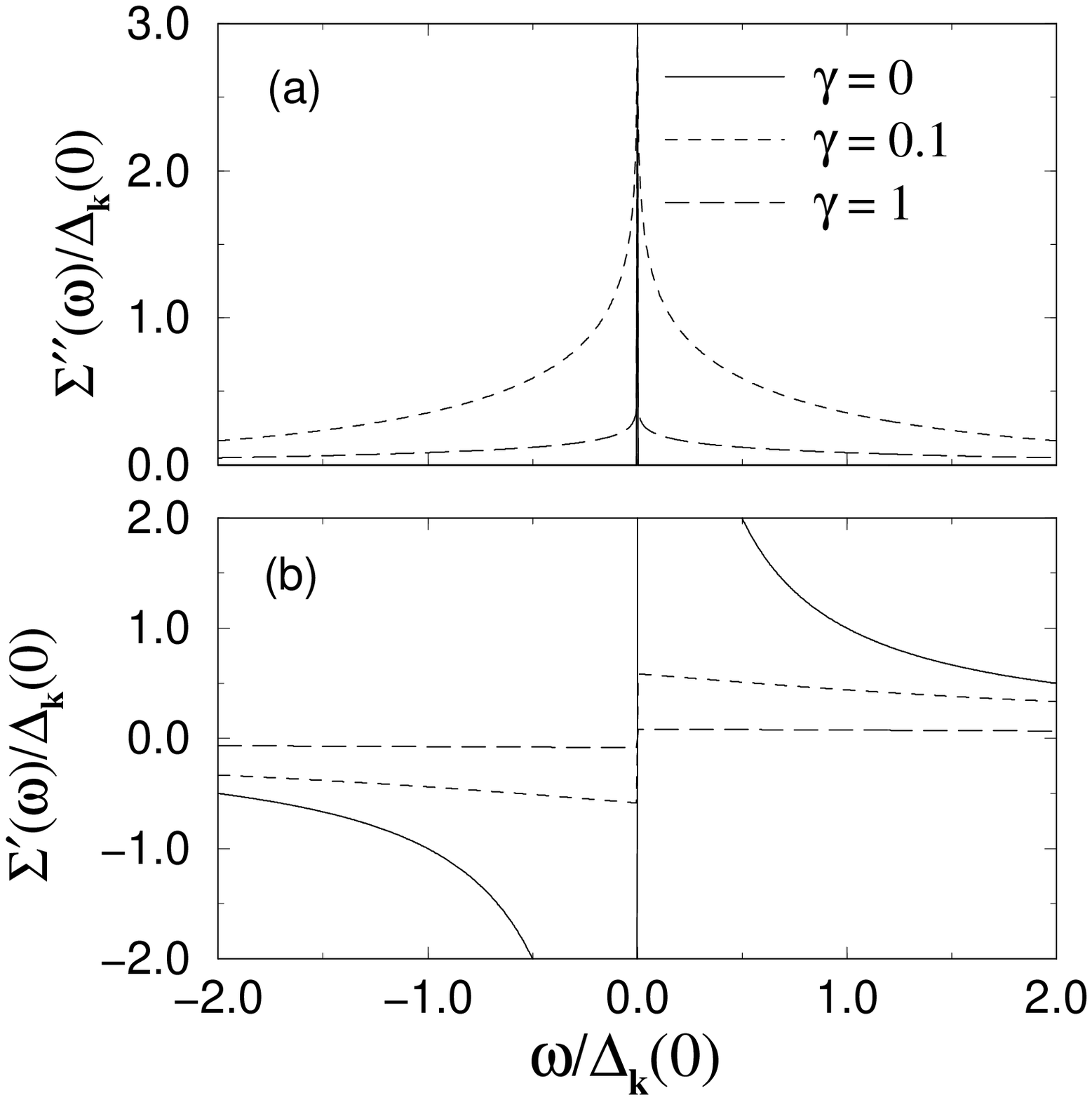}
}
\vspace{0.5cm}
\caption{
Plot of (a) the imaginary part
of the self-energy $\Sigma ^{\prime \prime }\left( \omega ,{\mathbf k},\gamma
,T=T_{c}\left( \gamma \right) \right) $ and (b) the real part of the
self-energy $\Sigma ^{\prime }\left( \omega ,{\mathbf k},\gamma
,T=T_{c}\left( \gamma \right) \right) $ as functions of $\omega $, of three
different values of the anisotropy parameter $\gamma $, and the same value
of the ratio $\Delta \left( \mathbf{k}\right) /\pi v_{\mathbf{k}}=0.064$. Each
self-energy is plotted at the transition temperature corresponding to $\gamma $.
}
\label{ir_vf=5_g=0.01.1_t0=1}
\end{figure}

In order to fully understand the effect of dimensionality, we have to take a
closer look at the analytic properties of the self energy at small energies,
for different values of the anisotropy $\gamma $, at $T=T_{c}\left( \gamma
\right) $.
The imaginary and the real part of the self-energy are plotted in Fig. (\ref
{ir_vf=5_g=0.01.1_t0=1}.a) and (\ref{ir_vf=5_g=0.01.1_t0=1}.b) respectively
for the same three values of $\gamma $ used in Fig. (\ref
{sf_vf=5_g=0.01.1_t0=1}). In exactly two dimensions, in which case $%
T_{c}\rightarrow 0$, the imaginary part of the self energy is a
delta function, and the real part of the self-energy is given by $%
\Delta _{\mathbf{k}}^{2}\left( 0\right) /\omega $. On the other hand, in
case of an isotropic three dimensional superconductor, at $T=T_{c}\left(
\gamma =1\right) $, the imaginary part of the self energy only diverges
logarithmically at $\omega =0$, while the real part is discontinuous ($%
\Sigma ^{\prime }\left( \omega \right) \propto $ sign$\left( \omega \right) $%
) but finite and small at $\omega =0$.

For large, but finite anisotropy ($\gamma \ll 1$), a crossover frequency $%
\omega _{X}=\gamma v_{\mathbf{k}}q_{BZ}^{\left( z\right) }$ appears in the
frequency dependence of $\Sigma \left( \omega \right) $. For $\omega <\omega
_{X}$ the behavior of the self energy is similar to the isotropic 3D
behavior, while for $\omega >\omega _{X}$ it has essentially the 2D
character. More specifically, for $\omega \ll \omega _{X}$ the imaginary
part of the self energy $\Sigma ^{\prime \prime }\left( \omega ,{\mathbf k}%
,\gamma ,T=T_{c}\left( \gamma \right) \right) $ diverges logarithmically at $%
\omega =0$, while a Taylor series expansion of the real part of the self
energy yields:

\begin{equation}
\Sigma ^{\prime }\left( \omega ,{\mathbf k},\gamma ,T=T_{c}\left( \gamma
\right) \right) \simeq \frac{\Delta _{\mathbf{k}}^{2}\left( 0\right)}{%
\omega _{X}}\frac{T_{c}\left( \gamma \right) }{T_{0}}\left\{ \frac{\pi ^{2}}{%
4}\text{sign}\left( \omega \right) -\left( 1+\widetilde{\gamma }^{2}\right) 
\frac{q_{BZ}^{\left( z\right) }}{q_{c}^{\left( z\right) }}\frac{\omega }{%
\omega _{X}}\right\}  \label{T=TcAsymptotic3D}
\end{equation}
which consists of a finite discontinuity at $\omega =0$, to be contrasted
with the infinite discontinuity in 2D.

\begin{figure}
\centerline{
\epsfxsize=4.0in
\epsfbox{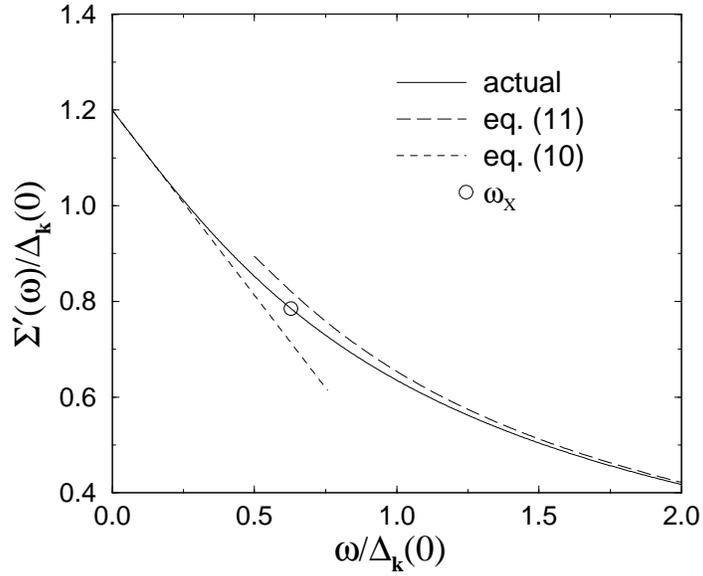}
}
\vspace{0.5cm}
\caption{
Plot of the real part of the
self energy $\Sigma ^{\prime }\left( \omega ,\mathbf{k}\right) $ solid line,
as a function of $\omega $, compared with the asymptotic expansions given in
eq. (\ref{T=TcAsymptotic2D}), long dashed line, and eq. (\ref
{T=TcAsymptotic3D}), dashed line. The value of $\Sigma ^{\prime }\left(
\omega \right) $ at the dimensional crossover frequency $\omega _{X}=\gamma
v_{\mathbf{k}}q_{BZ}^{(z)}$, is indicated by a circle. All curves are plotted 
for the same value of the anisotropy parameter $\gamma = 0.04$.
}
\label{re_vf=5_g=42_t0=1}
\end{figure}

This behavior is similar to that of the isotropic three dimensional
superconductor. As the anisotropy is increased ($\gamma $ $\rightarrow $ $0$%
) the jump in $\Sigma ^{\prime }$ at $\omega =0$ and the magnitude of the
slope in eq. (\ref{T=TcAsymptotic3D}) gradually increase to resemble the
divergence of $\Sigma ^{\prime }\left( \omega \rightarrow 0\right) $ in the
pure 2D case (see Fig. (\ref{ir_vf=5_g=0.01.1_t0=1}.b)). Now consider
excitation energies $\omega \gg \omega _{X}$ and $\gamma \ll 1$. In this
case the imaginary part of the self energy $\Sigma ^{\prime \prime }\left(
\omega ,{\mathbf k},\gamma ,T=T_{c}\left( \gamma \right) \right) $ crosses
over to a $\sim 1/\omega $ power law behavior, and the real part of the self
energy is well approximated by the following asymptotic expansion: 
\begin{equation}
\Sigma ^{\prime }\left( \omega ,{\mathbf k},\gamma ,T=T_{c}\left( \gamma
\right) \right) \simeq \frac{\Delta _{\mathbf{k}} ^{2}\left( T_{c}\left( \gamma
\right) \right) }{\omega }\left( 1+\frac{\frac{4}{3}+\log \left( \frac{%
\omega }{v_{\mathbf{k}}q_{c}}\right) -\frac{3}{2}\left( \frac{\omega }{v_{%
\mathbf{k}}q_{c}}\right) ^{2}}{\log \left( \frac{2}{\widetilde{\gamma }}%
\right) -\frac{1}{3}}\right)  \label{T=TcAsymptotic2D}
\end{equation}
This is the $\Delta ^{2}/\omega $ dependence expected for the two
dimensional case just before the SC phase transition $T\rightarrow T_{c}=0$,
plus corrections which vanish logarithmically in $\gamma $ as $\gamma
\rightarrow 0$. The two dimensionality of this behavior is made absolutely
clear by the observation that eq. (\ref{T=TcAsymptotic2D}) coincides with
the asymptotic expansion of the real part of the self energy calculated in
two dimensions at $T=T_{c}\left( \gamma \right) $. That is for $\omega \gg
\omega _{X}$, the third dimension is no longer relevant. In the next section
we shall see that this collapse of the self energy to its two dimensional
limit persists and actually extends to lower excitation energies as the
temperature is increased. Fig. (\ref{re_vf=5_g=42_t0=1}) exemplifies the
asymptotics just discussed for the real part of the self energy.

The role of the crossover energy $\omega _{X}=\gamma v_{\mathbf{k}%
}q_{BZ}^{\left( z\right) }$ in the experimental observation of the
pseudogap, such as in ARPES data, can be fully appreciated if $\omega _{X}$
is compared to the gap energy $\Delta _{\mathbf{k}}\left( 0\right) $. If $%
\Delta _{\mathbf{k}}\left( 0\right) \ll \omega _{X}(\mathbf{k})$, 
the properties of the
pseudogap will be 3D like. The position of the maxima in the spectral
function can be estimated by solving $\omega -\Sigma ^{\prime }\left( \omega
\right) =0$, where in this limit eq. (\ref{T=TcAsymptotic3D}) can be used
for $\Sigma ^{\prime }\left( \omega \right) $. The half width of the
pseudogap, $\Delta _{\mathrm{pg}}\left( {\mathbf k},\gamma ,T=T_{c}\left(
\gamma \right) \right) $, at the transition temperature is roughly
proportional to:

\begin{equation}
\frac{\Delta _{\mathrm{pg}}\left( {\mathbf k},\gamma ,T=T_{c}\left( \gamma
\right) \right) }{\Delta _{\mathbf{k}}\left( 0\right) }\propto \frac{\pi ^{2}%
}{4}\frac{T_{c}\left( \gamma \right) }{T_{0}}\frac{\Delta _{\mathbf{k}%
}\left( 0\right) }{\omega _{X}(\mathbf{k})}  \label{GapWidth3D}
\end{equation}
Therefore in case of a mildly anisotropic material, the pseudogap in the
spectral function may be completely unobservable, especially in the weak
coupling limit where $T_{0}\sim E_{F}$.

The situation changes considerably if high anisotropies are attained ($%
\gamma \ll 1$), and 
$\Delta _{\mathbf{k}}\left( 0\right) \gg \omega _{X}(\mathbf{k})$.
In this limit the spectral function is very well described by its two
dimensional limit except in a narrow range of excitation energies $\left|
\omega \right| \lesssim \omega _{X}\left( \gamma \right) $, and the width of
the pseudogap is expected to be of the order of $2\Delta _{\mathbf{k}}\left(
0\right) $.

Notice that the relation between $\Delta _{\mathbf{k}}\left( 0\right) $ and
$\omega _{X}=\gamma v_{\mathbf{k}}q_{BZ}^{\left( z\right) }$ is $\mathbf{k}$
dependent, that is it depends upon the particular point probed in the
Brillouin zone. In the case of high temperature superconductors, points on
the Fermi surface close to the zone diagonal have small values of $\Delta _{%
\mathbf{k}}\left( 0\right) $, and relatively high values of the Fermi
velocity. On the basis of this analysis for these points, the pseudogap
effect should not be observable, or should be very small even at $T=T_{c}$.
On the other hand, points on the Fermi surface close to the $\left( \pi
,0\right) $ point have large values of the gap, and small values of the
Fermi velocity, and here the pseudogap effect should be the strongest, as
indeed is experimentally observed\cite{Mike1}.

\smallskip

\section{T$\geq $T$_{c}\left( \gamma \right) $}

\begin{figure}
\centerline{
\epsfxsize=4.0in
\epsfbox{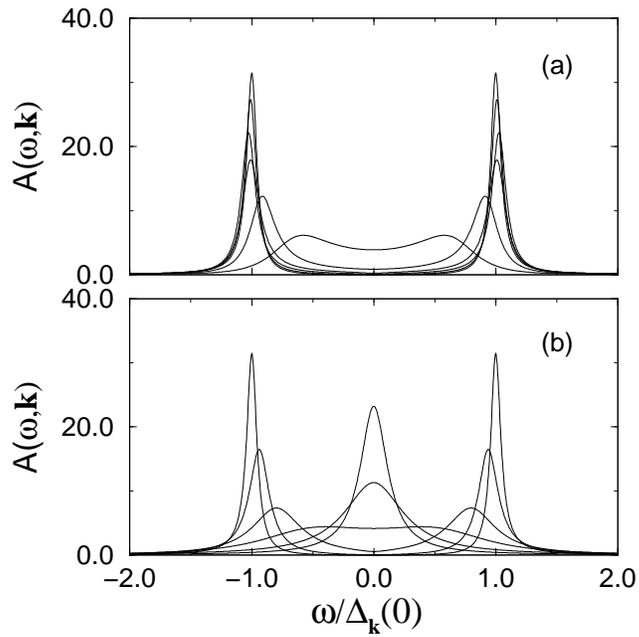}
}
\vspace{0.5cm}
\caption{
Plot of the spectral function%
\protect\cite{broadening} $A\left( \omega ,{\mathbf k},\gamma =0\right) $ in
2D as a function of $\omega $, at various temperatures ($T/T_{c} ^{MF} =
0.0, 0.1, 0.3, 0.5, 0.7, 0.9)$ for two different
values of the ratio $\Delta \left( \mathbf{k}\right) /v_{\mathbf{k}}$: (a) $%
\Delta \left( \mathbf{k}\right) /\pi v_{\mathbf{k}}=0.32$ and (b) $\Delta
\left( \mathbf{k}\right) /\pi v_{\mathbf{k}}=0.016$.
}
\label{sf_g=0_vf=1.20_t0=1}
\end{figure}

Let's now consider how the pseudogap evolves with temperature. For a given $%
\gamma $, as soon as the temperature is larger than $T_{c}\left( \gamma
\right) $, the singularities in the real and imaginary part of the self
energy at $\omega =0$ disappear. The singularity in $\Sigma ^{\prime \prime
}\left( \omega \right) $ is transformed into a peak of finite height, which
decreases and widens as the temperature increases, so that the spectral
function $A\left( \omega ,\mathbf{k}\right) $ does not vanish anymore at $%
\omega =0$, and the pseudogap begins to fill in. The discontinuity in $%
\Sigma ^{\prime }\left( \omega \right) $ is replaced by a line with a finite
slope given by:

\begin{equation}
\left. \frac{\partial \Sigma ^{\prime }\left( \omega ,{\mathbf k},\gamma
,T\right) }{\partial \omega }\right| _{\omega =0}=\frac{\Delta _{\mathbf{k}%
} ^{2}\left( 0\right) }{\left( v_{\mathbf{k}}q_{c}\right) ^{2}}\frac{T}{T_{0}}%
\frac{q_{c}^{\left( z\right) }}{q_{BZ}^{\left( z\right) }}\left[ \frac{%
1+\left( q_{c}\xi \right) ^{-2}}{\widetilde{\gamma }}\left( q_{c}\xi \right) 
\text{atan}\left( \widetilde{\gamma }q_{c}\xi \right) -1\right]
\label{ReSigSlope}
\end{equation}
As is clear from this equation, this slope is infinite at $T=T_{c}\left(
\gamma \right) $ (since $\xi \rightarrow \infty $), and decreases with
increasing temperature (since $\xi $ decreases). As long as this slope is
larger than one, the equation $\omega -\Sigma ^{\prime }\left( \omega ,%
{\mathbf k},\gamma ,T\right) =0$ has two non-zero solutions, and the spectral
function has two distinct peaks and a minimum at $\omega =0$. When $\left(
\partial \Sigma ^{\prime }/\partial \omega \right) _{\omega =0}<1$, the
equation $\omega -\Sigma ^{\prime }\left( \omega ,{\mathbf k},\gamma
,T\right) =0$ has only the solution $\omega =0$. Due to the maximum in $%
\Sigma ^{\prime \prime }\left( \omega ,{\mathbf k},\gamma ,T\right) $ at $%
\omega =0$, this may still correspond to a minimum in $A\left( \omega =0,%
\mathbf{k}\right) $, but the pseudogap is quickly suppressed as $\left(
\partial \Sigma ^{\prime }/\partial \omega \right) _{\omega =0}$ is further
reduced from $1$. Figures (\ref{sf_g=0_vf=1.20_t0=1}.a) and (\ref
{sf_g=0_vf=1.20_t0=1}.b) are examples of the temperature evolution of the
spectral function provided by this model.

The dimensional crossover energy, and its relation with the gap amplitude $%
\Delta _{\mathbf{k}}\left( 0\right) $, still determine the behavior of the
self energy and hence of the spectral function of the material. We have
already noticed in the previous section that, for $\gamma \ll 1$ and $\omega
\gg \omega _{X}$, the self energy of the anisotropic superconductor at the
transition temperature $T_{c}\left( \gamma \right) $ coincides with that of
the two-dimensional one, and therefore if $\Delta _{\mathbf{k}}\left(
0\right) \gg \omega _{X}(\mathbf{k})$ the values of the spectral function 
in these two
cases coincide except in a narrow range of excitation energies $\omega
\lesssim \omega _{X}$. This behavior persists as the temperature is
increased above $T_{c}\left( \gamma \right) $. This is because the
correlation length in the third dimension, $\xi _{z}$, decreases much faster
(by a factor of $1/\gamma $) than the in plane correlation length $\xi $,
and eventually becomes smaller than the interplane separation $a_{z}$,
thus reducing the material to a stack of nearly independent two dimensional
planes. Therefore, at temperatures high enough that $\xi _{z}\ll a_{z}$,
while still $\xi \gg $ $a$, the relation: 
\begin{equation}
\Sigma \left( \omega ,{\mathbf k,}\gamma ,T\right) \simeq \Sigma _{2D}\left(
\omega ,{\mathbf k},T\right)  \label{gamma==2d}
\end{equation}
is valid for all excitation energies and not only for $\omega \gg \omega
_{X} $ as we found at $T=T_{c}\left( \gamma \right) $. Therefore if $\Delta
_{\mathbf{k}}\left( 0\right) \gg \omega _{X}(\mathbf{k})$, 
the temperature evolution of
the pseudogap in the spectral function of an anisotropic material is
entirely two dimensional.

For $\Delta _{\mathbf{k}}\left( 0\right) <\omega _{X}(\mathbf{k})$, 
there is no such
general relation as eq. (\ref{gamma==2d}). The pseudogap is characterized by
a small separation of the broad maxima in the spectral function, at least
for small to intermediate coupling, and a small temperature range for which
these maxima are distinct.

\begin{figure}
\centerline{
\epsfxsize=4.0in
\epsfbox{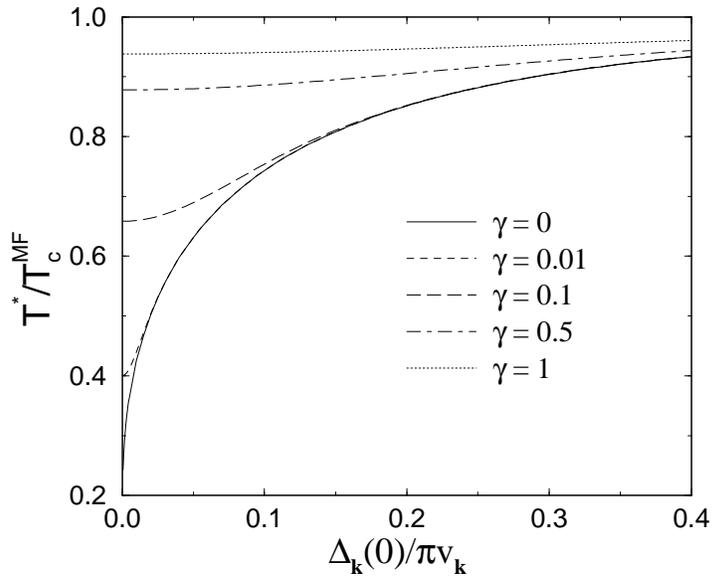}
}
\vspace{0.5cm}
\caption{
Plot of the estimated
temperature $T^{*}$ for the closing of the pseudogap, as a function of the
ratio $\Delta \left( \mathbf{k}\right) /\pi v_{\mathbf{k}}$ for various
values of the anisotropy parameter $\gamma $.
}
\label{tstr_t0=1._g=0.001.01.05.1}
\end{figure}

In order to estimate the temperature $T^{*}$ at which the pseudogap closes,
we can use the condition for having non-zero solutions of $\omega -\Sigma
^{\prime }\left( \omega ,{\mathbf k},\gamma ,T\right) =0$, which ceases to be
valid at a temperature $T^{*}$ such that:

\begin{equation}
\left. \frac{\partial \Sigma ^{\prime }\left( \omega ,{\mathbf k},\gamma
,T^{*}\right) }{\partial \omega }\right| _{\omega =0}\approx 1.
\label{T*Condition}
\end{equation}
Figure (\ref{tstr_t0=1._g=0.001.01.05.1}) is a plot of $%
T^{*}/T_{MF}$ as a function of the ratio $\Delta _{\mathbf{k}}\left(
0\right) /v_{\mathbf{k}}$. Clearly in the limit $\Delta _{\mathbf{k}}\left(
0\right) /v_{\mathbf{k}}\rightarrow 0$, $T^{*}\rightarrow T_{c}\left(
\gamma \right)$, which means that close to a node the pseudogap in
the spectral function can open only in an extremely small temperature range.
As the ratio $\Delta _{\mathbf{k}}\left( 0\right) /v_{\mathbf{k}}$
increases, $T^{*}/T_{c}^{MF}$ increases for any value of $\gamma $, but the
range of the pseudogap effect in the spectral function $\left(
T^{*}-T_{c}\left( \gamma \right) \right) $ increases more rapidly for higher
anisotropies, so that it is larger the larger the anisotropy. Notice also
that the curves at high anisotropy (small $\gamma $) collapse on top of the $%
\gamma =0$ curve as soon as $\Delta _{\mathbf{k}}\left( 0\right) \gtrsim
\omega _{X}\left( \gamma, \mathbf{k} \right) $. This is a consequence of 
the dimensional
crossover just discussed (eq. (\ref{gamma==2d})): the pseudogap in the
spectral function will disappear at the same temperature for anisotropic
materials with different $\gamma $ as long as $\Delta _{\mathbf{k}}\left(
0\right) \gg \omega _{X}\left( \gamma, \mathbf{k} \right) $.

It is important to notice once again that the condition $\Delta _{\mathbf{k}%
}\left( 0\right) \gg \omega _{X}\left( \gamma, \mathbf{k} \right) $ 
is $\mathbf{k}$
dependent. This is in qualitative agreement with the experimentally observed
phenomenon of the Fermi arcs\cite{Mike1}. It is apparent from photoemission
experiments that $T^{*}$ depends on which points of the Fermi surface are
probed. The general trend is that the pseudogap near the $\left( 0,0\right) $%
-$\left( \pi ,\pi \right) $ direction closes at a lower temperature than
that near the $\left( \pi ,0\right) $ one. This is well captured by eq. (\ref
{T*Condition}) and (\ref{ReSigSlope}). For points on the Fermi surface close
to the zone diagonal, the Fermi velocity $v_{\mathbf{k}}$ is at its maximum,
while at the same time, if a d-wave symmetry for the order parameter is
assumed, $\Delta _{\mathbf{k}}\left( 0\right) \ $is small. This makes the
ratio $\Delta _{\mathbf{k}}\left( 0\right) /v_{\mathbf{k}}$ small and
therefore $T^{*}$ small (see Fig. (\ref{tstr_t0=1._g=0.001.01.05.1})).
Conversely, close to $\left( \pi ,0\right) $ the Fermi velocity $v_{\mathbf{k%
}}$ is at its minimum, and for a d-wave order parameter, $\Delta _{\mathbf{k}%
}\left( 0\right) $ is large, which makes $T^{*}$ large.

\smallskip

Finally we want to comment on another peculiarity of the temperature
evolution of the pseudogap which is in good qualitative agreement with the
experimentally observed phenomenon of the closing up versus filling in
behavior of the pseudogap at different points in $\mathbf{k}$-space. Fig. (%
\ref{sf_g=0_vf=1.20_t0=1}.a) and (\ref{sf_g=0_vf=1.20_t0=1}.b) exemplify the
contrast in the temperature evolution of the spectral function for larger
and smaller values of the $\mathbf{k}$ dependent ratio $\Delta _{\mathbf{k}%
}\left( 0\right) /v_{\mathbf{k}}$, which resembles quite closely the ARPES
experimental data. For points on the Fermi surface close to $\left( \pi
,0\right) $, the ratio $\Delta _{\mathbf{k}}\left( 0\right) /v_{\mathbf{k}}$
is relatively large, and the spectral function evolves like in Fig. (\ref
{sf_g=0_vf=1.20_t0=1}.a), that is the peaks seem rather immobile until the
temperature is close to $T_{c}^{MF}$ and the pseudogap appears to fill up.
On the other hand, for points close to the zone diagonal, the ratio $\Delta
_{\mathbf{k}}\left( 0\right) /v_{\mathbf{k}}$ tends to be small and the
behavior is more like that of Fig. (\ref{sf_g=0_vf=1.20_t0=1}.b), that is
the pseudogap appears to close in rapidly with temperature. However in Fig. (%
\ref{sf_g=0_vf=1.20_t0=1}.a) another interesting feature is evident: the
separation of the peaks as the temperature is increased increases a little
at first before it starts to collapse. It can be shown that this effect is a
consequence of our assumption that $\Delta _{\mathbf{k}}\left( T\right) $ is
given by the BCS expression, which is almost constant at temperatures $T\ll
T_{c}^{MF}$. A relaxation of this assumption could eliminate the effect.
Further investigation of this effect would require a more accurate model
for $\Delta _{\mathbf{k}}\left( T\right) $, which is beyond the scope of
this paper.

\smallskip

\section{Conclusions}

We have demonstrated that classical pairing fluctuations give rise to a
pseudogap phenomenon in the limit of small to intermediate coupling
strength. However, in both 2D and 3D, the dimensionality plays a crucial role
in the strength, shape, and temperature dependence of the pseudogap. In
three dimensional isotropic or weakly anisotropic systems, the effect is
very weak. The magnitude of the pseudogap just above the transition
temperature is reduced from its $T=0$ value at least by a factor of the
order of $\Delta _{\mathbf{k}}\left( 0\right) /\gamma v_{\mathbf{k}%
}q_{BZ}^{\left( z\right) }$, which in the case of an isotropic
superconductor would be of the order of $\Delta /E_{F}$, and its temperature
range is quite small. Large values of the coupling strength are therefore
necessary to observe a pseudogap in a material with small anisotropy. On the
other hand, in two dimensions, or for relatively high anisotropy, the
pseudogap is a sizeable effect even for small to intermediate coupling
strength: the magnitude of the pseudogap just above the transition
temperature is almost the same as its $T=0$ value, while its temperature
range is also considerably increased, $T^{*}$ being of the order of $%
T_{c}^{MF}$ provided that $\xi _{0}\sim 1$. The dimensional crossover
between these two qualitatively different regimes is defined by $\gamma \sim
\Delta _{\mathbf{k}}\left( 0\right) /v_{\mathbf{k}}q_{BZ}^{\left( z\right) }$%
. For values of $\gamma $ significantly smaller than $\Delta _{\mathbf{k}%
}\left( 0\right) /v_{\mathbf{k}}q_{BZ}^{\left( z\right) }$, the pseudogap is
describable in terms of the two dimensional limit, while for $\gamma $
essentially larger than $\Delta _{\mathbf{k}}\left( 0\right) /v_{\mathbf{k}%
}q_{BZ}^{\left( z\right) }$, the anisotropic three dimensional limit is
proper. Furthermore, the in plane anisotropy plays a crucial role in
determining the temperature range of the pseudogap through the ratio $\Delta
_{\mathbf{k}}\left( 0\right) /v_{\mathbf{k}}$, so that close to the 
$\left( \pi ,0\right) $ point, the Fermi velocity is small (Van Hove point) 
and $T^{*}$
is large, while for points on the Fermi surface for which the ratio $\Delta
_{\mathbf{k}}\left( 0\right) /v_{\mathbf{k}}$ is small, the pseudogap
has a smaller range.

The parameters characterizing the underdoped cuprates, namely high
anisotropies, large superconducting gap amplitudes, and small correlation
lengths, put these materials in the 2D limit, and our results are in rather
good qualitative agreement with experiment.

\smallskip

\section{Acknowledgments}
This work was supported by the U. S. Dept. of Energy, Basic Energy 
Sciences, under contract W-31-109-ENG-38, and the National Science 
Foundation DMR 91-20000 through the Science and Technology Center 
for Superconductivity.

\newpage

\smallskip

\begin{thebibliography}{99}
\bibitem{gtilda}  Although this expression is strictly speaking valid only
in the weak coupling limit, it can actually be extended up to intermediate
coupling by replacing the bare interaction with a renormalized one. See for
example\cite{Vilk}.
\bibitem{CutoffProcedure}  In obtaining this equation we have used a
``soft'' cutoff procedure, which ensures that $\chi \left( 0,\mathbf{q}%
\right) $ approaches $0$ continuously at $q=q_{c}$, in order to avoid the
spurious singularities that a sharp cutoff would give rise to in the self
energy. For this particular calculation, we have replaced $\chi _{0}$ with $%
\chi _{0}U\left( \mathbf{q}\right) $with $U\left( \mathbf{q}\right)
=1-q_{c}^{-2}\left( q_{x}^{2}+q_{x}^{2}+\gamma ^{2}q_{x}^{2}\right) $.
\bibitem{broadening}  In this and other plots of the spectral function, a
broadening factor $\Gamma _{1}=0.1\times \Delta _{\mathbf{k}}\left( 0\right) 
$ has being added to the imaginary part of the self energy, in order to make
the sharp peaks visible.
\end{thebibliography}
\end{document}